\documentclass[preprint,pre,showpacs]{revtex4}
\date{\today}
\usepackage{graphicx}
\usepackage[pdfpagemode=None,pdfstartview=FitH]{hyperref}
\usepackage{epstopdf}

\def\L{{\bf L}}
\def\r{{\bf r}}
\def\p{{\bf p}}
\def\k{{\bf k}}
\def\csch{ { \text{csch}}}
\def\coth{ { \text{coth}}}

\begin{document}

\title{Exact Solution to Ideal Chain with Fixed Angular Momentum}

\author{J. M. Deutsch}
\affiliation{ Department of Physics, University of California, Santa Cruz, CA 95064.}

\begin{abstract}
The statistical mechanics of a non-interacting polymer chain in the limit of a
large number of monomers 
is considered when the total angular momentum, $\L$, is fixed. The radius
of gyration for a ring polymer in this situation is derived exactly in
closed form by functional integration techniques. Even when $L = 0$ the
radius of gyration differs from that of a random walk by a prefactor of
order unity. The dependence on $L$ is discussed qualitatively and the
large $L$ limit can be understood by physical arguments.
\end{abstract}


\maketitle

The statistical properties of polymers have been
the subject of intensive research for many decades~\cite{degennes}.
However these efforts have been almost entirely confined to  polymers in liquids
or solids, while in contrast,
their properties in a vacuum have received little
attention. Until recently there were not clear physical realizations of
such situations, but now with recent developments in the mass spectrometry
of proteins~\cite{Hillenkamp} and the search for long
hydrocarbon molecules in interstellar media~\cite{NewCarbonChains}, such systems are now of experimental
interest. The author~\cite{DeutschPolyVac} recently  considered such systems theoretically and by
means of computer simulation, and the purpose  of
this paper is to present an exact derivation for the radius of gyration of
a polymer in a vacuum with conserved angular momentum. 

In reality angular
momentum is weakly broken by interaction with thermal electromagnetic
radiation~\cite{DeutschPolyVac} but it is still important to understand
the case of conservation laws properly in order to understand these more
complicated effects. Also intra-chain interactions are not considered,
that is this is the case of an ``ideal" chain~\cite{degennes}.  What we
will find is at first sight rather surprising, that a polymer chain with
conserved total energy $E$, total linear momentum $\p_{tot} = 0$, and total momentum $\L$, has
a radius of gyration that depends strongly on $L$, so even when $L=0$,
the radius of gyration differs significantly from that of an ideal chain without
this restriction.

The statistical mechanics of a general classical system of $N$
particles with constant total energy, momentum, and angular momentum
with coordinates $\{\r_i\}$ and momenta $\{{\bf p}_i\}$ has been
considered previously by Laliena~\cite{laliena}. They were considered
to be interacting via a general potential $\Phi$. In the case under
consideration here, this microcanonical formulation can be expressed in 
a canonical ensemble in the limit of large $N$, after which it can be
converted to a functional integral.  This can then be evaluated to obtain
the radius of gyration as a function of angular momentum $\L$.

Laliena~\cite{laliena} has shown that the conservation of
linear momentum does not effect answers obtained in the microcanonical ensemble
with conservation of angular momentum  enforced~\cite{footnote:linearP}. So we will write down the
volume of phase space with $\L$ and energy $E$ kept constant.
\begin{equation}
W(E,L,N) = C \int \delta(E - K  - \Phi) \delta^{(3)}(\L - \sum_i \r_i\times\p_i) \delta^{(3)}(\r_{cm})\left(\prod_{i=1}^N d^3 r_i d^3 p_i\right)
\label{eq:MicroVolume}
\end{equation}

$C$ is a constant here that involves $N$ and $\hbar$ and is of no consequence for the
purposes here.  $K$ is the kinetic energy $\sum_i p_i^2/2m$, with $m$ the mass
of each monomer, and here we are taking them all to be equal. The center of mass
$\r_{cm}$ also must be conserved and is set to $0$.
We use the Fourier representation of the $\delta$ functions to write this as
\begin{equation}
W(E,L,N) \propto  \int_C d\lambda e^{\lambda E}  \int 
  e^{-\lambda( K  + \Phi)}  \delta^{(3)}(\L - \sum_i \r_i\times\p_i) \delta^{(3)}(\r_{cm}) \left(\prod_{i=1}^N d^3 r_i d^3 p_i\right)
\label{eq:MicroVolumeContour}
\end{equation}
As shown by Lax~\cite{Lax}, for most purposes, as discussed below,
the contour of integration can be deformed in the complex
$\lambda$ plane using the method of steepest descents. The three conditions,
(a) that a saddle point exists, (b) that an observable not be of order $\exp(const. N)$,
and (c) that there be no singularity in the observable in the neighborhood of the
saddle point. Conditions (b) and (c) are first obtained through the canonical
ensemble and then tested to see if they are satisfied. Condition (a), that a
saddle point exists, is satisfied because we can find a relationship between
the energy and the temperature. In the case of an athermal system, say of
rigid links, this would just be that $E \propto 1/\beta$, where $\beta$ is
the value of $\lambda$ at the saddle point. Condition (b) is satisfied for
the quantity of interest here, the average radius of gyration. Condition (c)
is also satisfied because we will see that the average radius of gyration
is smooth function of the temperature for T for finite $T = 1/\beta$.
Thus we can drop the integration over $\lambda$ and replace $\lambda$ by
the inverse temperature, $\beta$, and consider the partition function
$Z$ instead instead of the phase space volume integral (which is simply
related to the entropy.)
\begin{equation}
Z(\beta,L,N) \propto  \int d^3 k  \int e^{i\k \cdot \L}  
  e^{-\beta( K  + \Phi)} e^{- i \k \cdot \sum_i \r_i\times\p_i} \delta^{(3)}(\r_{cm}) \left(\prod_{i=1}^N d^3 r_i d^3 p_i\right)
\label{eq:Canonicalrp}
\end{equation}
Integrating over the $\p_i$'s we obtain
\begin{equation}
Z(\beta,L,N) \propto  \int d^3 k  e^{i\k \cdot \L} \int e^{-\frac{1}{2\beta} \k\cdot I \cdot \k  }  
  e^{-\beta \Phi} \prod_{i=1}^N d^3 r_i  \equiv  \int d^3 k  e^{i\k \cdot \L} \zeta(\beta,\k)
\label{eq:Canonicalr}
\end{equation}
where $I$ is the moment of inertia tensor for the particles
\begin{equation}
I_{\alpha \gamma} = m \sum_{i=1}^N ({r^\nu}_i {r^\nu}_i \delta_{\alpha \gamma} - {r^\alpha}_i {r^\gamma}_i)
\end{equation}
where $\alpha$ and $\gamma$ label the coordinates ($1,2,3$), and the Einstein summation
convention has been used for $\nu$. In the last equality of Eq. \ref{eq:Canonicalr}
we have introduced the function $\zeta(\beta,\k)$. Note that this cannot depend on the direction
of $\k$ but only its magnitude, if $\Phi$ only involves isotropic central potentials. 
Therefore we can take $\k$ to be along the $z$ axis: $\k = k {\hat z}$, and write
\begin{equation}
\zeta(\beta,k) = \int e^{-\frac{m k^2}{2\beta} \sum_i (x_i^2+y_i^2) -\beta \Phi}  \delta^{(3)}(\r_{cm})\prod_{i=1}^N d^3 r_i 
\label{eq:zeta}
\end{equation}
Because of the radial dependence of $\zeta$ on $\k$ we can also perform the $\k$ angular integrals in
Eq. \ref{eq:Canonicalr}, rewriting the $\k$ integration in spherical coordinates, obtaining
\begin{equation}
Z(\beta,L,N) = \frac{c}{L} \int_0^\infty  k \sin(k L) \zeta(\beta,k) d k
\label{eq:ZandZeta}
\end{equation}
where $c$ is a constant that plays no role in the subsequent analysis.

The potential is taken to be that of an ideal Gaussian chain with step length
$l$ and a ring topology. 
\begin{equation}
\beta \Phi_0 = \frac{3}{2 l^2} \left( \sum_{i=1}^{N-1} |\r_{i+1}-\r_i)|^2 + |\r_N-\r_1|^2\right)
\end{equation}
By the central limit theorem, many models of polymer chains will all
give the same results for most quantities of interest, if the overall radius of
gyration is $\ll N$.

In order to calculate the radius of gyration, one can add an additional
potential with a parameter $\epsilon$ 
\begin{equation}
\beta \Phi = \beta \Phi_0 +   \epsilon l \sum_{i=1}^N |\r_i|^2 
\end{equation}
so that the average radius of gyration can be written as
\begin{equation}
R_g^2 = \langle \frac{1}{N}\sum_{i=1}^N |\r_i|^2 \rangle =  -\frac{1}{N l} \frac{\partial \ln Z}{\partial \epsilon}{\Big |}_{\epsilon=0}
\label{eq:RgDef}
\end{equation}
The integration in Eq. \ref{eq:zeta} is Gaussian 
and we can now take the usual limit to turn this into a functional integral
\begin{equation}
\zeta(\beta,k) = \int e^{-\int_0^{N l}  (\frac{T m k^2}{2 l} + \epsilon)(x^2(s) + y^2(s))+ \epsilon z^2 + \frac{3}{2l} |\dot{\r}|^2 ds}  
    \delta^{(3)}(\r_{cm})\delta \r(s)
\end{equation}
The functional integration in the $x$, $y$, and $z$ directions decouple and the
$x$ and $y$ functional integrals are identical.
Each one of these three integrals is of the form of the partition function of a one dimensional
quantum harmonic oscillators at finite temperature except for the restriction on the center of mass.
If we consider the Euclidean time action for a quantum harmonic oscillators of
mass $M$ at inverse temperature $\beta_o$
\begin{equation}
S = \int_0^{\beta_o} \frac{M}{2}\left(\dot{x}^2 + \omega_0^2 x^2\right) dt
\end{equation}
then the partition function
\begin{equation}
Z_o = \int e^{-S} \delta x(t) \propto \frac{1}{2 \sinh(\frac{\beta_o \omega_0}{2})}
\label{eq:Zo}
\end{equation}
with periodic boundary conditions on the paths $x(0) = x(\beta_o)$.
This is true because of the general formula relating the partition function to
the Euclidean path integral over times ranging from $0$ to the inverse
temperature~\cite{FeynmanStatMech}. The partition function, Eq. \ref{eq:Zo}, can also be derived 
in a less elegant but more direct manner by writing all paths
in terms of a Fourier expansion, which then decouples the integrals, and forms
an infinite product over all modes.
This latter approach is useful in the present application because we
have the additional restriction on the path integral that the zero mode
should not be integrated over as a consequence of the restriction on the center
of mass. Using the Fourier decomposition approach, we can easily incorporate
this restriction, by not including the zero mode in the
product. This amounts to multiplying Eq. \ref{eq:Zo} by $\omega_0$

Rescaling variables so that the angular momentum $L' \equiv L\sqrt{12}/(Nl\sqrt{m T})$ 
and using Eq. \ref{eq:RgDef} gives
\begin{equation}
\frac{{R_g}^2}{N l^2} = \frac{1}{36} \frac{\int_0^\infty (k(-6 + k^2  + 6 k \coth(k)) \text{csch}(k)^2 \sin(k L')) dk}
{\int_0^\infty k^3 \text{csch}(k)^2 \sin (k L') dk}
\end{equation}

Both the numerator and denominator can be computed in closed form using
contour integration. This gives the final result
\begin{equation}
\frac{{R_g}^2}{N l^2} = \frac{
2 L' \left(3+\pi ^2\right)+L' \left(\pi ^2 - 6\right)\
\text{cosh}[L' \pi ]+3 \left(L'^2 -1\right)
\pi  \text{sinh}[L' \pi ]
}
{
36\pi  (2 L' \pi\
+L' \pi  \text{cosh}[L' \pi ]-3 \text{sinh}[L' \pi ])
}
\end{equation}

\begin{figure}[t]
\begin{center}
\includegraphics[width=0.9\hsize]{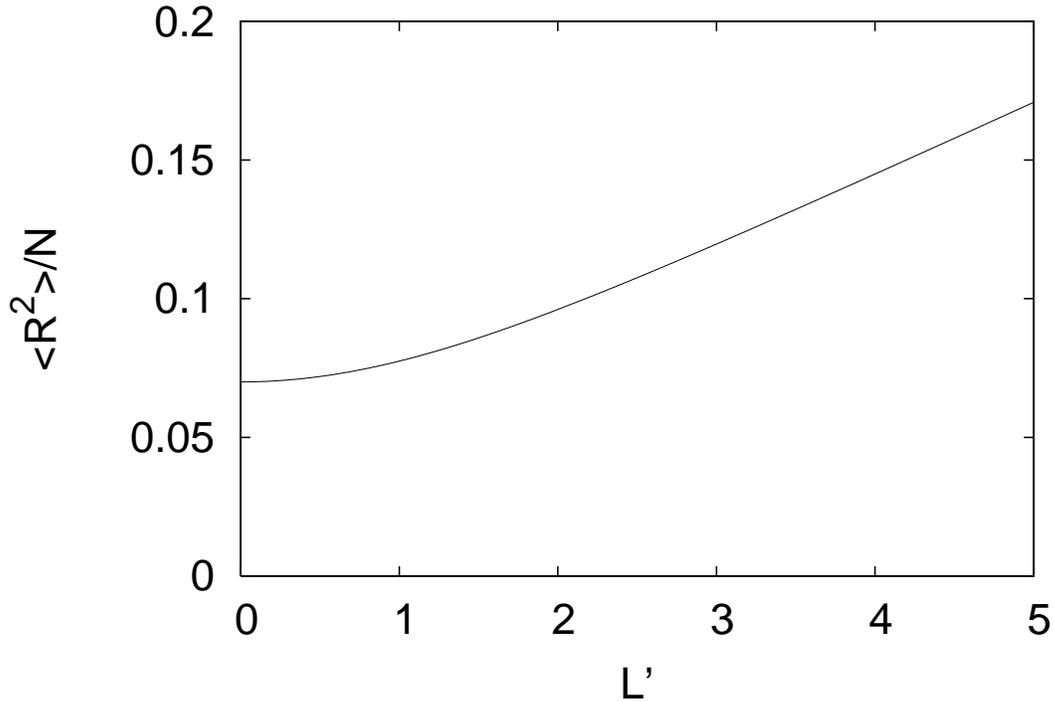}
\caption{ 
The radius of gyration versus rescaled angular momentum
for an ideal ring chain.
}
\label{fig:rg}
\end{center}
\end{figure}

A plot of this equation is displayed in Fig. \ref{fig:rg}.

For $L'=0$ this reduces to $R_g^2/(N l^2) = (1 + 15/\pi^2)/36 \approx 0.07$. For
a ring without angular momentum conservation, $R_g^2/(N l^2) = (1/12)  \approx 0.083$
which means that the restriction to $L = 0$ causes the rings to be smaller
relative to the case where the angular momentum can take on any value.

In the opposite limit of large $L'$, $R_g^2/(N l^2) \rightarrow L'/{12 \pi}$.
Note that in terms of $L$, $R_g^2 \rightarrow L l/(\pi \sqrt{12 T m})$ independent
of chain length $N$. To understand this behavior, we consider high $L$
configurations, where we expect that a typical configuration of the ring will be close to a circle
rotating rapidly. The approximate free energy contains a kinetic energy and an
elastic term
\begin{equation}
F = \frac{L^2}{2 I} + \frac{k}{2} C^2
\label{eq:LargeF}
\end{equation}
Where $C = 2\pi R$ is the circumference, and $k$ is the entropic elastic spring
coefficient $k = 3T/(N l^2)$. The moment of inertia is approximately $I = m N R^2$.
Minimizing with respect to $R^2$ this gives the above result. It is not
surprising that this result is exact because in the large $L$ limit, we expect that this
circular configuration will become dominant.

The temperature in this model at a given energy is determined in the usual
way, by requiring that the average energy in the canonical ensemble is
equal to the microcanonical energy. However in this case, the radius
of gyration is a very sensitive function of $L$. To see this, note
that changes occur on a scale $L' \sim 1$, or $L \sim N l \sqrt{mT}$.
Because $I$ is typically $\sim (N m) R_g^2 \sim (N m) (N l^2) \sim
N^2 l^2 m$ then $L \sim N l \sqrt{m T}$.  The order of $L^2/2I$  is
therefore $\sim T$. This means that a change of order one degree of
freedom changes $R_g^2/(N l^2)$ by a number of order unity, which has
a negligible effect on the temperature but a large effect on the radius
of gyration. So for fixed $L'$ as $N \rightarrow \infty$, we see that in
the canonical ensemble, the effect of the angular momentum constraint on
the energy is a fraction of order $1/N$.  This means that when the limit
$N\rightarrow \infty$ is taken with $L'$ fixed, the relation between the
temperature and energy can be obtained as it would for a polymer without
angular momentum conservation, and thus will not have any dependence on
$L'$. This is also seen more rigorously by computing the exact dependence
of the partition function on $L'$, which is done below.

It is useful to calculate the probability density for finding the polymer
with a  particular value of total (rescaled) angular momentum $L'$. This should be important
in the case where there is a dilute gas of such polymers. It is also important for
a single chain for long times, since the angular momentum is changed by the weak coupling to electromagnetic
black body radiation~\cite{DeutschPolyVac}. In this case the probability
density function $P(L')$ is proportional to the partition function $Z(\beta,L',N)$.
The normalization requirement  is that
\begin{equation}
\int_0^\infty  P(L') 4\pi {L'}^2 d{L'} = 1
\end{equation}

The normalization is straightforward to calculate
using Eq. \ref{eq:ZandZeta} and integrating over $L'$ first. The $L'$
integration requires evaluating
\begin{equation}
\int_0^\infty  L' \sin(kL') dL' = - \int_0^\infty \frac{d \cos(kL')}{d k} dL' = -\pi \delta'(k)
\end{equation}
and using this, the integral over $k$ is now easily accomplished 
\begin{equation}
\int_0^\infty  Z(\beta, L', N) 4\pi {L'}^2 dL =  c\pi \frac{\partial (k \zeta(\beta,k))}{\partial k}\Big|_{k=0} 
\label{eq:norm}
\end{equation}

Z was evaluated previously in the process of calculating the radius of gyration
and is proportional to 
\begin{equation}
\int_0^{\infty} k^3 \csch(k)^2 \sin(k L') dk / L'
\end{equation}

Evaluating this integral and including the correct normalization using Eq. \ref{eq:norm}
yields
\begin{equation}
P(L') = \frac{\pi^3 \csch(\frac{L' \pi}{2})^4 ( 2 L' \pi + L' \pi \cosh(L'\pi) - 3\sinh(L'\pi))}{16 L' \pi^2}
\end{equation}

\begin{figure}[t]
\begin{center}
\includegraphics[width=0.9\hsize]{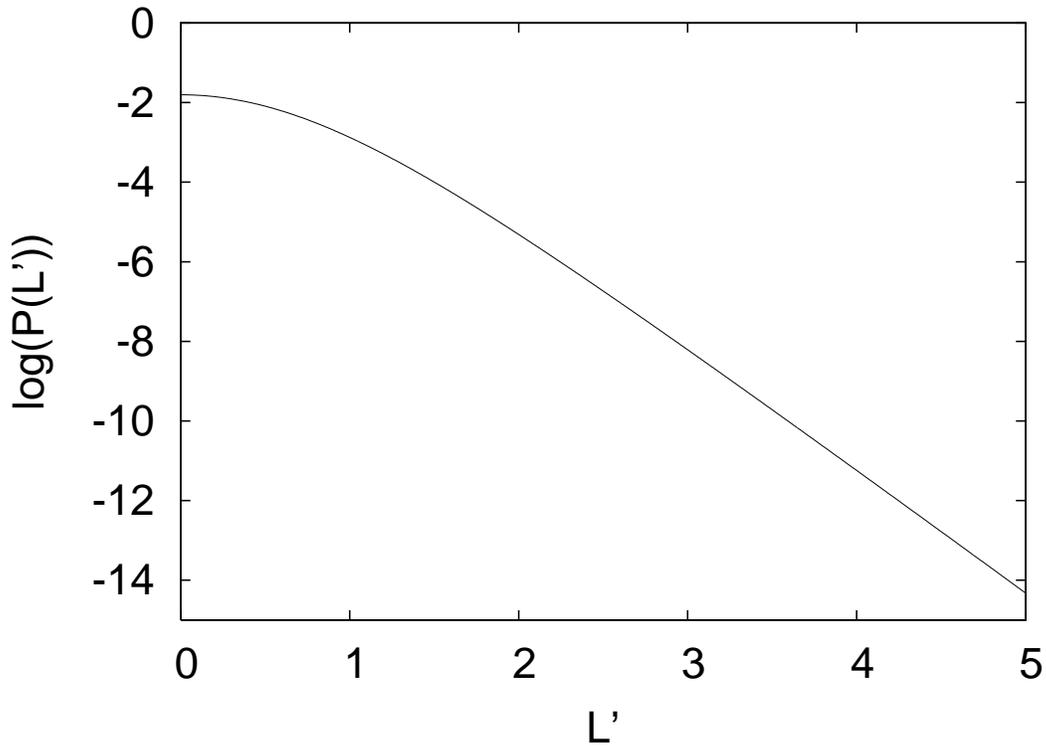}
\caption{ 
The probability density function for finding a chain
in thermal equilibrium with rescaled angular momentum $L'$
for an ideal ring chain.
}
\label{fig:pdf}
\end{center}
\end{figure}

$\ln(P(L'))$ versus $L'$ is  plotted in Fig. \ref{fig:pdf}.
Because of the non-constant value of the moment of inertia for the chain,
this distribution is decidedly non-gaussian. In the large $L'$ limit,
the slope of this curve approaches a constant with a slope of $-\pi$.
This is in agreement with the minimization argument for large $L'$
given under Eq.  \ref{eq:LargeF}.

It is interesting to compare the extreme sensitivity of this system to
restrictions in angular momentum, with what would be expected
in other kinds of systems. The system considered here is essentially
one dimensional in that interactions are only from nearest neighbor
monomers. In, for example, a membrane or a three dimensional gel,
the system is of higher dimension. In such two or three dimensional
systems, a perturbation that changes the free energy by ${\cal O}(k_B T)$
is expected to only effect averages by microscopic amounts. However for
polymers, it has a much larger effect.  For example, a force pulling the
ends of a polymer costing $k_B T$ of energy will increase its radius
of gyration by a multiplicative constant of order unity.  Because a
polymer chain with the restriction $L=0$, reduces the number of degrees
of freedom by at least one, it should affect the free energy by ${\cal O}
(k_B T)$.  So by the above argument, this is expected to make non-trivial
changes to its statistics, unlike higher dimensional systems. So even in
the case $L=0$, the non-interacting polymer is no longer an uncorrelated
random walks.

The author wishes to thank Peter Young for useful discussions.


\begin{thebibliography}{}
\bibitem{degennes} P.G. de Gennes ``Scaling Concepts in Polymer Physics" Cornell University Press (1985).
\bibitem{Hillenkamp} F. Hillenkamp (Editor), J. Peter-Katalinic (eds.), 
``MALDI MS: A Practical Guide to Instrumentation, Methods and Applications'' Wiley, (2007).
\bibitem{NewCarbonChains} P. Thadeus, M.C. McCarthy, M.J. Travers, C.A. Gottlieb, and W. Chen,
Faraday Discuss., {\bf 109}, 121 (1998).
\bibitem{DeutschPolyVac} J.M. Deutsch, Phys. Rev. Lett. {\bf 99}, 238301 (2007).
\bibitem{laliena} V. Laliena, Phys. Rev. E {\bf 59}, 4786 (1999).
\bibitem{footnote:linearP} Alternatively in the derivation that follows, the extra
$\delta$ function constraints on the total linear momentum can be included
in their Fourier representation. It is easily seen by completing the square,
that this constraint can be absorbed into redefinitions of the $\p_i$'s and
three extra gaussian integrations, which have no affect on the chain's
statistics.
\bibitem{Lax} M. Lax, Phys. Rev. {\bf 97} 1419, (1955).
\bibitem{FeynmanStatMech} R.P. Feynman, ``Statistical Mechanics: A Set of
Lectures'' Basic Books (1998), Page 72.
\end{thebibliography}
\end{document}